
\def\AP#1{{\sl Ann.\ Phys.\ (N.Y.) {\bf #1}}}

\def\CMP#1{{\sl Comm. Math. Phys. {\bf #1}}}

\def\JETPL#1{{\sl JETP Lett.\ {\bf #1}}}

\def\PLB#1{{\sl Phys.\ Lett.\ {\bf #1B}}}

\def\PRD#1{{\sl Phys.\ Rev.\   {\bf D #1}}}


\def\nxl{\hfill\break}




\def\a{\alpha}

\def\b{\beta}
\def\g{\gamma}

\def\e{\epsilon}

\def\l{\lambda}

\def\s{\sigma}

\def\t{\theta}

\def\te{\vartheta}


\def\o{\over}

\def\bold#1{\setbox0=\hbox{$#1$}
     \kern-.025em\copy0\kern-\wd0
     \kern.05em\copy0\kern-\wd0
     \kern-.025em\raise.0433em\box0 }
\def\lowmp{\lower.11em\hbox{${\scriptstyle\mp}$}}

\def\frac#1#2{{\textstyle{
 #1 \over #2 }}}                            


\def\1{{\rm 1 \!\!\, l}}                        
%

%
%


\hyphenation{Di-par-ti-men-to}
\hyphenation{na-me-ly}
\hyphenation{al-go-ri-thm}
\hyphenation{pre-ci-sion}
\hyphenation{cal-cu-la-ted}

\input phyzzx
\Pubnum={$\rm PAR\; LPTHE\; 93/53
         \qquad {\rm November \; 1993}$}
\date={}
\titlepage
\title{{\bf THE GENERAL SOLUTION OF THE 2-D
SIGMA MODEL STRINGY BLACK
HOLE AND THE MASSLESS COMPLEX SINE-GORDON MODEL.}}
\author{ H.J. de Vega }
\address{ Laboratoire de Physique Th\'eorique et Hautes Energies, Paris
     \foot{Laboratoire Associ\'e au CNRS UA 280 \nxl
      Postal address: \nxl
           L.P.T.H.E., Tour 16, $1^{\rm er}$ \'etage,
        Universit\'e Paris VI,\nxl
        4, Place Jussieu, 75252, Paris cedex 05, FRANCE }}
\author{  J. Ram\'{\i}rez Mittelbrunn, M. Ram\'on Medrano }
\address{ Departamento de F\'{\i}sica Te\'orica, Madrid
\foot{Postal address: \nxl Facultad de Ciencias F\'{\i}sicas, Universidad
Complutense,  Ciudad Universtaria, \nxl E-28040, Madrid, ESPA\~NA.} }
\author{N. S\'anchez}
\address{Observatoire de Paris, Section de Meudon, Demirm
\foot{Laboratoire Associ\'e au CNRS UA 336, Observatoire de Meudon et
\'Ecole Normale Sup\'erieure.\nxl  Postal address: \nxl
DEMIRM, Observatoire de Paris. Section de Meudon, 92195 MEUDON
Principal Cedex, FRANCE.}}
\endpage
\vfil
\abstract
The exact general solution for a sigma model having the $2-d$ stringy black
hole (SBH) as internal manifold is found in closed form. We also give
the exact solution for the massless complex Sine-Gordon (MCSG) model.
Both, models and their solutions are related by analytic continuation.
The solution is expressed in terms of four arbitrary functions of one variable.
\endpage
\sequentialequations
\REF\lure{F. Lund, T. Regge, \PRD{14}, 1524, (1976)}
\REF\lund{F. Lund, \AP{115}, 251 (1978)}
\REF\pohl{K. Pohlmeyer, \CMP{46}, 207 (1976), \nxl
B.S. Getmanov, \JETPL{25}, 119, (1977)}
\REF\quan{H.J. de Vega, J.M. Maillet, \PRD{28}, 1441 (1983)}
\REF\edw{E. Witten, \PRD{44}, 314 (1991)}
\REF\cuar{H.J. de Vega, J. Ramirez, M. Ram\'on Medrano, N. S\'anchez, PAR LPTHE
93/14}
\REF\klin{W. Klingenberg, A course in Differential Geometry, Springer
Verlag, New York 1978.}
\REF\grie{ I. Bakas and E. Kiritsis, \PLB{301}, 49 (1993), \nxl
I. Bakas, CERN preprints 7046 and 7047, 1993.}
\vskip 1cm
The two dimensional conformal field theory given by the Lagrangian
$$
{\cal L} = {1 \o 2} ~
 {{\partial_{a}U \partial^{a} V}
\o {1 - UV}}
\eqn\lagra
$$
where $U$ and $V$ are in general (complex) null type field coordinates, appears
in different and relevant physical contexts:

(i) Eq.\lagra\  is the
zero mass limit of the complex Sine-Gordon (CSG) model
[\lure,\lund,\pohl,\quan].
The CSG model is a relativistic massive theory which is integrable at the
classical and quantum levels. The CSG model can be obtained as a
reduction of the $O(4)$ sigma model [\pohl]. In addition, the
(massive) CSG model
describes the dynamics of a classical relativistic string coupled to a
constant external scalar field in flat $3+1$ dimensional space time
[\lure].

 (ii) Eq.\lagra\  describes a classical string on a two dimensional
stringy black hole background [\edw,\cuar]. The massless field theory
defined by \lagra\ exhibits a $W_{\infty}$ symmetry [\grie ].

The case (i) is described by eq.(1) with the reality condition
$$
U = V^{*} \eqn\estre
$$
For the case (ii) the reality conditions are
$$
U = U^{*} \qquad , \qquad V = V^{*}
\eqn\dobes
$$
Thus, both theories, the stringy black hole (SBH) and the massless
complex Sine-Gordon (MCSG) are related by analytic continuation.

The aim of this letter is to obtain the exact general solution for
these theories in a closed form. The equations of motion are
$$\eqalign{
\partial_{a}\left({{\partial^{a}V}\o{1-UV}}\right) =&~
V \, {{\partial_{a}U\partial^{a}V}\o{(1-UV)^2}} \cr
\partial_{a}\left({{\partial^{a}U}\o{1-UV}}\right) =&~
U \, {{\partial_{a}U\partial^{a}V}\o{(1-UV)^2}} \cr}
\eqn\movi
$$
Introducing $(r,t)$ coordinates given by
$$
U = \sinh r ~ e^t \qquad , \qquad V = - \sinh r ~ e^{-t}
\eqn\coor
$$
we can summarize the general solution of eqs.\movi\ as follows
$$
\cosh^2r(\s,\tau)={1 \o 2}\left[1 + \vec{A}(\s+\tau). \vec{B}(\s-\tau)
\right]
\eqn\erre
$$
and
$$\eqalign{
t(\s,\tau)=&{1 \o 2}\int_{a}^{\s+\tau}\left({{\e_{\a \b \g}
A'^{\a}A^{\b}B^{\g}}\o{\vec{A}.\vec{B}-1}}\right)(x_+,b)~dx_+ \cr
-&{1 \o 2}\int_{b}^{\s-\tau}\left({{\e_{\a \b \g}
B'^{\a}A^{\b}B^{\g}}\o{\vec{A}.\vec{B}-1}}\right)(\s+\tau,x_{-})~dx_-
+c \cr}
\eqn\te
$$
Here, $\vec{A}(\s+\tau)$  and $\vec{B}(\s-\tau)$ are arbitrary
 three-component vectors lying on the one
sheeted hyperboloids
$$\eqalign{
\vec{A}.\vec{A}=& -(A_0)^2 +(A_1)^2 +(A_2)^2 = 1 \cr
\vec{B}.\vec{B}=& -(B_0)^2 +(B_1)^2 +(B_2)^2 = 1 \cr}
\eqn\hip
$$
($\tau$ and $\s$ are the time and space coordinates, respectively)  $'$ stands
for
the derivative with respect to the argument, and $a, b $ and $c$ are arbitrary
real constants.

In the $(r,t)$ variables, eqs.\coor\ , the lagrangian density, eq. \lagra\ ,
and
the equations of motion, eq. \movi\ , read respectively $$
{\cal L} = {1 \o 2} \left[ (\partial_{a}r)^2 -
\tanh^2r~(\partial_{a}t)^2 \right] \eqn\lsbh
$$
$$\eqalign{
\partial_{a}\left(\tanh^2r~ \partial^{a}t\right)&=0 \cr
\partial_{a}\partial^{a}r + {{\sinh~r}\o{\cosh^3
r}}(\partial_{a}t)^2&=0\cr} \eqn\sbh
$$
For the SBH theory, the reality conditions, eq. \dobes\  amount to
$$\eqalign{
{\rm (i)}~  t^* =& t~ {\rm and}~ r^* = r~ {\rm for}~ U\,V < 0 \cr
{\rm (ii)}~  (t - i\pi/2)^* =& t - i\pi/2 , ~{\rm and}~
 r^* = - r ~Ê{\rm for}~ 0 <U\,V < 1 \cr}
\eqn\reco
$$
Regions (i) and (ii) are, respectively, the outside of the black hole
and the region inside the
horizon ($(r,t)$ are Schwarzschild coordinates).

Eqs.\sbh\ are the equations of motion of a sigma model defined on a
stringy black hole manifold as internal space. This is the coset
$SL(2,R)/U(1)$ with metric
$$
ds^2 = -\tanh^2 r~(dt)^2 + (dr)^2
$$
For the MCSG theory the following coordinates are usually introduced
[\lund,\pohl]
$$
U = e^{i\l}~\cos\t \qquad ,\qquad V= e^{-i\l}~\cos\t
\eqn\ccsg
$$
In terms of $\l$ and $\t$, the lagrangian density and the equations of
motion read
$$
{\cal L} = -{1 \o 2} \left[ (\partial_{a}\t)^2
+ \cot^2\t ~ (\partial_{a}\l )^2 \right]
\eqn\llt
$$
$$\eqalign{
\partial_{a}\left(\cot^2\t~ \partial^{a}\l\right) & =0 \cr
\partial_{a}\partial^{a}\t +
 {{\cos\t}\o{\sin^3\t}}(\partial_{a}\l)^2 & = 0 \cr}
\eqn\lth
$$
The reality condion \estre\ translates simply into $\l^* = \l$
 and $\t^* = \t$ . From eqs.\coor\
and eqs. \ccsg\  the fields $r$ and $t$ are related to the fields
$\t$  and $\l$ by the analytic transformation
$$
r= i(\t - \pi/2) \qquad, \qquad t=i(\l + \pi/2)
\eqn\prol
$$
We consider now the SBH theory and proceed to obtain the general
solution of equations \sbh\ . For that purpose we shall use the
construction of ref.[\lure] in which the equations for a relativistic one
dimensional extended object coupled to a massless scalar field lead
to the equations of the massive Sine-Gordon system as integrability
conditions. In ref. [\lure], particular solutions of the massive
Sine-Gordon model in $1+1$ dimensions were used to construct
 solutions of strings coupled to
an external field in a $3+1$ dimensional Minkowski
space-time. Instead, we shall work in the opposite sense: we shall
start with a free string in $2+2$ flat space-time to produce the
general solution of the equations for the SBH theory.
(A free string in $3+1$ dimensional Minkowski spacetime produces
the general solution of the MCSG theory).

The world sheet surface $X^{\mu}(\s,\tau)=({\tilde X}^0,X^0,X^1,X^2)(\s,\tau)$
 of a free string moving in a four
dimensional flat space with metric diag$(-,-,+,+)$  satisfies the
wave equation
$$
\left({{\partial^2}\o{\partial\tau^2}} - {{\partial^2}\o{\partial\s^2}}\right)
X^{\mu} = 0 
\eqn\dalu
$$
and the constraints
$$\eqalign{
\left({{\partial X^\mu}\o{\partial\tau}}\right)^2+
\left({{\partial X^\mu}\o{\partial\s}}\right)^2 &= 0 \cr
{{\partial X^\mu}\o{\partial\s}}.{{\partial X_\mu}\o{\partial\tau}}=0 \cr}
\eqn\vinu
$$
Choosing the gauge ${\tilde X}^0 = \tau$ and projecting the world sheet
 $X^{\mu}(\s,\tau)$ over the $1+2$
hyperplane $(X^0,X^1,X^2)$ we obtain a surface $M$ given by
 $X^{\a}(\s,\tau), (\a =0, 1, 2)$ which satisfies the
following equations
$$
\left({{\partial^2}\o{\partial\tau^2}} - {{\partial^2}\o{\partial\s^2}}\right)
X^{\a} = 0
\eqn\dald
$$
$$\eqalign{
\left({{\partial X^\a}\o{\partial\tau}}\right)^2+
\left({{\partial X^\a}\o{\partial\s}}\right)^2 &= 1 \cr
{{\partial X^\a}\o{\partial\s}}.{{\partial X^\a}\o{\partial\tau}}=0 \cr}
\eqn\vind
$$
whose general solution is
$$
X^{\a}(\s,\tau)= \phi^{\a}(\s+\tau) + \psi^{\a}(\s-\tau)
\eqn\sold
$$
Here $\phi^{\a}$ and $\psi^{\a}$ are arbitrary functions satisfying
$$
(\phi'^{\a})^2=(\psi'^{\a})^2= 1/4
\eqn\norm
$$
Then, introducing the notation,
$$
A^{\a}\equiv2~\phi'^{\a}\qquad,\qquad B^{\a}\equiv 2 ~\psi'^{\a}
\eqn\defab
$$
the class of surfaces $M$ that satisfy equations \dald\ and \vind\ can
be described by two $2+1$ real vectors  of arbitrary
functions, $A^{\a}(\s+\tau)$ and
$ B^{\a}(\s-\tau)$, lying on the one-sheeted hyperboloids \hip\ .

On the other hand, the metric on $M$,
induced by the flat metric $(-,-,+,+)$ takes the form
$$
ds^2 = (\partial_{\s}X^\mu)^2 ~d\s^2 + 2 ~ \partial_{\s}X^\mu
\partial_{\tau}X^\mu ~d\s d\tau +  (\partial_{\tau}X^\mu)^2 ~d\tau^2
$$
Taking into account eqs.\vind\ , this metric
 can be parametrized as
$$
ds^2 = -\sinh^2r~d\tau^2 + \cosh^2r~d\s^2
\eqn\metr
$$
and the metric function $\cosh^2r$ for the surface $M$ takes the form
 of eq.\erre\ .

Here $r$ can be either real or purely imaginary, as follows from the
conditions (i)-(ii).

We would like to remark here that the black hole metric \lagra\
$$
 G_{ab}  = \pmatrix{ -\tanh^2r &0 \cr 0 &1 \cr} \quad
\eqn\bhm
$$
and the world-sheet metric \metr\
$$
g_{ab}=\pmatrix{ -\sinh^2r &0 \cr 0 &\cosh^2r \cr} \quad
\eqn\wsm
$$
differ by an obvious conformal factor $\cosh^2 r$, which coincides with
the exponential of the dilaton field in this case [\edw ]:
$$
\Phi(r) = \ln \cosh^2r
$$
Now, as it is well known in classical differential geometry, the
surface $M$ satisfies the pseudoeuclidean version of the
Gauss-Weingarten equations [\klin]
$$
\partial_a\partial_b X^{\a} = \Gamma^c_{ab}\;\partial_c X^{\a} +
 h_{ab} \;N^{\a}
\eqn\gw
$$
determining the embedding of the surface $M$ in the flat spacetime.
Here $\Gamma^c_{ab}$ are the Christoffel symbols corresponding to the metric
\metr\ , on the surface $M$, $N^{\a}$ is the
unit normal to the surface $M$ given by
$$
N^{\a} = { {n_{\a}} \o {\sqrt{n_{\a}n^{\a}}}}\;  ,~
{\rm where}~~ n_{\a} = \e_{\a\b\g}\,
\partial_{\tau} X^{\b}\, \partial_{\s} X^{\g}
\eqn\ene
$$
and $h_{ab}$ is the extrinsic curvature or second fundamental form
$$
h_{ab} = - \partial_a N^{\a}\partial_b X_{\a}=N^{\a}
\partial_a \partial_b X_{\a}
\eqn\hac
$$
The integrability conditions for eqs.\gw\  are the Gauss equations
$$
R_{abcd} = h_{ac} h_{bd}-h_{ad}h_{bc}
$$
where $R_{abcd}$  is the Riemann curvature tensor on $M$,
and the Codazzi-Mainardi equations
$$
\Gamma^c_{ab} h_{dc} - \Gamma^c_{da}h_{bc}+\partial_d h_{ab}-
\partial_b h_{da} = 0
$$
In two dimensions there is only one independent Gauss equation and
two independent Codazzi-Mainardi equations. For a surface that
satisfies \dald\  and \vind\  the metric and the second fundamental
form can be written as matrices of the form
$$
g_{ab}=\pmatrix{ -E &0 \cr 0 &G \cr} \quad
\eqn\gab
$$
$$
h_{ab}=\pmatrix{ H & L  \cr L &H \cr} \quad
\eqn\hab
$$
and then the Gauss and Codazzi-Mainardi equations reduce to
$$\eqalign{
\partial_{\s}^2E -\partial_{\tau}^2G +{1\o 2 }
\partial_{\tau}G~\partial_{\tau}\ln EG -{1\o 2 } \partial_{\s}E~
\partial_{\s}\ln EG &= 2~( H^2 - L^2 ) \cr
H\left({1 \o E} - {1 \o G} \right)\partial_{\s}E+ L \partial_{\tau}\ln
[G/E] + 2~\partial_{\tau} L - 2~\partial_{\s}H &= 0 \cr
H\left({1 \o E} - {1 \o G}
\right)\partial_{\tau}G + L \partial_{\s}\ln [G/E] + 2~\partial_{\tau}H
- 2~\partial_{\s}L &= 0 \cr } \eqn\gcm
$$
In the particular case of our metric \metr\ ,  equations \gcm\
become
$$\eqalign{
\sinh r ~\cosh r~\partial^a \partial_a r& = H^2 - L^2 \cr
\partial_{\tau}[L \coth r] &=\partial_{\s}[ H \coth r] \cr
\partial_{\tau}[H \tanh r] &=\partial_{\s}[ L \tanh r] \cr}
\eqn\gcmd
$$
Let us see now how the SBH sigma model equations of motion
\sbh\  actually arise from eqs.\gcmd\ .
We deduce from the second equation \gcmd\
 that there exists locally a field $t(\s,\tau)$ such that
$$
{{\partial t} \o {\partial \tau}} = H \coth r \qquad , \qquad
{{\partial t} \o {\partial \s}} = L \coth r
\eqn\ult
$$
which inserted in the other two equations \gcmd\  yields
 equations \sbh\ .

In eqs.\gcmd\  we used the metric \metr\  for which
$ f(r) \equiv {E \o G} = \tanh^2 r $,
but notice that we could consider a world-sheet metric with an
arbitrary $f(r)$ and $ G - E = 1 $. Then the Codazzi-Mainardi
equations would take the form
$$\eqalign{
\partial_{\tau}\left[{{L}\o {\sqrt{f(r)}}}\right] &
=\partial_{\s}\left[{H\o {\sqrt{f(r)}}}\right]  \cr
\partial_{\tau}[H\sqrt{f(r)} ] &=\partial_{\s}[ L \sqrt{f(r)}] \cr}
$$
However, these equations together with the Gauss equation \gcm\  would
derive from a lagrangian density
$$
{\cal L} = {1 \o 2} \left[ (\partial_{a}r)^2 -
f(r)^2~(\partial_{a}t)^2 \right] \eqn\lf
$$
if and only if $ f(r) = \tanh^2 r $ or  $ f(r) = \coth^2 r $ .

We summarize now what we have done from equations \metr\  to \ult\  :
the fields $r$ and $t$ have been identified as particular
combinations of the functions entering the first and second
fundamental forms of the surface $M$. Furthermore, the
 SBH sigma model equations of motion \sbh\ , satisfied by the fields $r$ and
$t$,
have been also identified with part of the Gauss and Codazzi-Mainardi equations
for $M$. Then, since we know the class of surfaces $M$ satisfying \dald\
and \vind\ , we also know the general solutions for the SBH sigma model
equations of motion. More precisely, the unit normal $N^{\a}$, eq. \ene\ , and
the second fundamental form functions, $H$ and $L$ \hac\ , have the
following expressions in terms of the arbitrary functions
 $\vec{A}(\s+\tau)$  and $\vec{B}(\s-\tau)$ lying on the hyperboloids \hip\
$$\eqalign{
N_{\a} =& {1 \o {\sqrt{ ({\vec A} . {\vec B} )^2 - 1}}}~
\e_{\a \b \g} A^{\b}B^{\g} \cr
H =&  {1 \o {2 \sqrt{ ({\vec A} . {\vec B} )^2 - 1}}}~
\e_{\a \b \g} A^{\b}B^{\g}(A'^{\a} + B'^{\a})  \cr
L =&  {1 \o {2 \sqrt{ ({\vec A} . {\vec B} )^2 - 1}}}~
\e_{\a \b \g} A^{\b}B^{\g}(A'^{\a} - B'^{\a})  \cr }
$$
These equations inserted in \ult\  yield
$$
\partial_+ t = {1 \o 2} {1 \o {{\vec A} . {\vec B} - 1}}~
\e_{\a \b \g} A^{\b}B^{\g} A'^{\a} \quad , \quad
\partial_- t = -{1 \o 2} {1 \o {{\vec A} . {\vec B} - 1}}~
\e_{\a \b \g} A^{\b}B^{\g} B'^{\a}
\eqn\dert
$$
where we have defined $ \s_{\pm} = \s \pm \tau$.

Thus, eq.\te\  together with eq.\erre\  (which yield $t(\s,\tau)$
and $\cosh^2r(\s,\tau)$ provide closed
formulae for the general solutions of the sigma model SBH equations.
Furthermore, recalling eq.\hip\  we see that the general solutions for
the fields $r(\s,\tau)$ and $t(\s,\tau)$  depend on the four arbitrary
real valued functions of one variable :
$\a(\s_+), \phi(\s_+), \b(\s_-)$ and $\psi(\s_-)$,
 which are introduced through the parametrization
$$\eqalign{
(A^0,A^1,A^2) = (\sinh \a, \cosh \a \cos \phi,\cosh \a \sin \phi )
\cr
(B^0,B^1,B^2) = (\sinh \b, \cosh \b \cos \psi,\cosh \b \sin \psi )
\cr }
\eqn\para
$$
Nevertheless, since the equations of motion are conformal invariant,
one of the functions depending on $\s_+$ and
one of the functions depending on $\s_-$ can be chosen arbitrarily.

It is easy to check that these solutions describe massless lumps
which scatter with unit S-matrix, as expected for a conformal
invariant theory in two dimensions.

The energy-momentum tensor for the SBH sigma model is
$$
{\cal T}_{\pm \pm} = {1 \o 2} \left[ (\partial_{\pm}r)^2 -
\tanh^2r ~ (\partial_{\pm}t)^2 \right]
\eqn\tmm
$$
It is conserved
$$
\partial_{\mp}{\cal T}_{\pm \pm} = 0
$$
and for the general solution, \erre\ and \te\ , it takes the form
$$
{\cal T}_{++} = - { {A'^2(\s + \tau)}\o 8} ~~,~~
{\cal T}_{--} = - { {B'^2(\s - \tau)}\o 8}
$$
Concerning the reality conditions (i) and (ii), eqs.\reco\ ,  some comments are
in order. When ${\vec A} . {\vec B} \geq 1$, according to equations
\erre\  and \te\ , we can take $r$ and
$t$ to be real and we are in the region outside the black hole. On
the other hand, when  $0 \leq {\vec A} . {\vec B} \leq 1$
 equation \erre\  tells us that $r$ is pure
imaginary, while equation \te\  allows us to have $(t - i \pi/2)^*
= t - i \pi/2 $, by choosing the
arbitrary constant to be $i\pi/2$, and we are then in the region inside the
horizon. Finally, ${\vec A} . {\vec B} \leq 0$
 corresponds to the solutions in the unphysical
region $ UV > 1 $.  We also note that the singularity for
 $ {\vec A} . {\vec B} =  1$
 in equations \dert\
gives the expected discontinuity for the Schwarzschild time $t$ at
the crossing of the horizon, in such a way that two different
integration origins must be taken in formula \te\  at both sides of
the horizon.

The constraint equations of the free string in $2+2$ flat spacetime
define (in the $X^0 = \tau$ gauge) the sigma model manifold
$SL(2,R)/U(1)$ which is the SBH manifold.
If we want to consider strings propagating in this background,
besides the equations of motion \sbh\ we should impose the
string constraints
$$
{\cal T}_{\pm \pm} = 0
$$
By imposing these constraints, the general solution given by eqs.
\erre\ - \te\ can be reduced to the string solution presented in ref.[\cuar].

Let us turn now to the MCSG case. As we mentioned above, the
solution of the MCSG equations \lth\  can be obtained by analytic
continuation of the SBH sigma model solution. In fact, we analytically
continue the formulae \hip\ , \erre\  and \te\ for complex $A^{\a}$ and
$B^{\a}$.  We perform the substitutions
 $$
A^0 = i A^3 \qquad , \qquad B^0 = i B^3
$$
and introduce the notation
$$
{\vec A} =  (A^1,A^2,A^3) ,~{\vec B} =  (B^1,B^2,B^3).
$$
Then, we obtain
$$\eqalign{
\cosh^2 r &= {1 \o 2} (1 -  {\vec A} . {\vec B}) \cr
\partial_+t &= {1 \o 2} {i \o {{\vec A} . {\vec B} - 1}}
({\vec A} \wedge {\vec B}).{\vec A'} \cr
\partial_-t &= -{1 \o 2} {i \o {{\vec A} . {\vec B} - 1}}
({\vec A} \wedge {\vec B}).{\vec B'} \cr}
\eqn\bas
$$
$$
{\vec A}^2 = 1 = {\vec B}^2
\eqn\hipd
$$
By using equations \prol\ these equations can be rewritten as
$$\eqalign{
\sin^2\t &= {1 \o 2} (1 +  {\vec A} . {\vec B}) \cr
\partial_+\l &= {1 \o 2} {1 \o {{\vec A} . {\vec B} - 1}}
({\vec A} \wedge {\vec B}).{\vec A'} \cr
\partial_-\l &= -{1 \o 2} {1 \o {{\vec A} . {\vec B} - 1}}
({\vec A} \wedge {\vec B}).{\vec B'} \cr
{\vec A}^2 &= 1 = {\vec B}^2 \cr}
\eqn\smcs
$$
Now, for real ${\vec A}$ and ${\vec B}$
 ($\t$ and $\l$ are hence real), equations \smcs\
 provide the general solution for the MCSG theory.
Furthermore, condition \hipd\ tells us that this solution depends on
the four arbitrary real valued functions
${\hat \a}(\s_+), {\hat \phi}(\s_+), {\hat \b}(\s_-)$ and
${\hat \psi}(\s_-)$, through the parametrization
$$\eqalign{
{\vec A} = (\sin {\hat \a}\cos {\hat\phi}, \sin {\hat\a}
\sin{\hat\phi} , \cos {\hat \a} ) \cr
{\vec B} = (\sin  {\hat \b}\cos {\hat\phi},
 \sin{\hat \b} \sin  {\hat \psi},\cos  {\hat \b})
\cr }
\eqn\pard
$$
We also note that the parametrization \pard\ is easily related to
its pseudoeuclidean counter-part by
$$
\a = i (  {\hat \a} - \pi/2 ) ~,~  {\hat \phi} = \phi ~,~
\b = i (  {\hat \b} - \pi/2 ) ~,~  {\hat \psi} = \psi
$$
We would like to point out that the general solution for the MCSG
theory can alternatively be obtained by starting with a free string
in a $3+1$ dimensional Minkowski space-time and in the gauge
$ X^0 = \tau $. Then,
one considers the projection $M$ of the world-sheet over the three
dimensional euclidean space, and one identifies eqs. \smcs\  as
part of the Gauss and Codazzi-Mainardi equations for the surface $M$.

{\bf ACKNOWLEDGMENTS}

 The authors acknowledge the french-spanish scientific
 cooperation for partial support sponsored by Ministerio de
Educaci\'on y Ciencia (DGICYT) (Spain), Minist\`ere des
Affaires Etrang\`eres and Minist\`ere de L'Education Nationale
(France).

J.R.M. and M.R.M. would like to acknowledge the hospitality at LPTHE
(Universit\'e Pierre et Marie Curie) and at DEMIRM (Observatoire de
Paris-Meudon).

\refout

\bye